\documentclass[conference]{IEEEtran}
\pdfoutput=1 
\IEEEoverridecommandlockouts
\usepackage{cite}
\usepackage{amsmath,amssymb,amsfonts}
\usepackage{algorithmic}
\usepackage{graphicx}
\usepackage{textcomp}
\usepackage{xcolor}
\def\BibTeX{{\rm B\kern-.05em{\sc i\kern-.025em b}\kern-.08em
    T\kern-.1667em\lower.7ex\hbox{E}\kern-.125emX}}

\usepackage{url}
\usepackage[pagebackref,breaklinks,colorlinks]{hyperref}

\usepackage{gensymb}
\usepackage{tikz,pgfplots,forest}
\usepackage{tikz-qtree}
\usepackage{multirow}
\usepackage{booktabs}
\usepackage{caption} 
\usepackage{tabularx}

\begin{document}

\title{Machine Learning Techniques for Source Localisation in Elastic Media\\}
% {\footnotesize \textsuperscript{*}Note: Sub-titles are not captured in Xplore and
% should not be used}
% \thanks{Identify applicable funding agency here. If none, delete this.}
% }

\author{\IEEEauthorblockN{Bansi Mandalia$^1$, Steve Greenwald$^1$, Simon Shaw$^2$, Gregory Slabaugh$^1$\\
$^1$Queen Mary University of London \\
$^2$Brunel University London}
% \IEEEauthorblockA{\textit{School Of Electronic Engineering and Computer Science} \\
% \textit{Queen Mary University of London}\\
% London, UK \\
% b.mandalia@se17.qmul.ac.uk}
% \and
% \IEEEauthorblockN{2\textsuperscript{nd} Greg Slabaugh}
% \IEEEauthorblockA{\textit{School Of Electronic Engineering and Computer Science} \\
% \textit{Queen Mary University of London}\\
% London, UK \\
% g.slabaugh@qmul.ac.uk}
% \and
% \IEEEauthorblockN{3\textsuperscript{rd} Steve Greenwald}
% \IEEEauthorblockA{\textit{
% Blizard Institute - Faculty of Medicine and Dentistry} \\
% \textit{Queen Mary University of London}\\
% London, UK \\
% s.e.greenwald@qmul.ac.uk}
% \and
% \IEEEauthorblockN{4\textsuperscript{th} Simon Shaw}
% \IEEEauthorblockA{\textit{College of Engineering, Design and Physical Sciences} \\
% \textit{Brunel University London}\\
% London, UK \\
% simon.shaw@brunel.ac.uk}
 }

\maketitle
\thispagestyle{plain}
\pagestyle{plain}

\begin{abstract}
Coronary Artery Disease (CAD) results from plaque deposit in a coronary artery. Early diagnosis is imperative, so a non-invasive detection method is being developed to identify acoustic signals caused by partial occlusions in the artery. The blood flow in the artery is disturbed and imposes oscillatory stresses on the artery wall. The deformations caused by the stresses can be detected at the chest surface. Therefore, by using data simulating these surface signals, which arise from randomly assigned source positions, machine learning (ML) can be utilised to predict the source of the occlusion. Seven ML algorithms were investigated, and the results from this study found that an ensemble model combining k-Nearest Neighbours and Random Forest had the best performance. The metrics used to evaluate this was the mean squared error and Euclidean distance. 
\end{abstract}

\begin{IEEEkeywords}
machine learning, regression, non-invasive diagnosis, coronary artery disease, source localisation 
\end{IEEEkeywords}

\section{Introduction}
\begin{figure}[!ht]
\centerline{\includegraphics[width=0.35\textwidth]{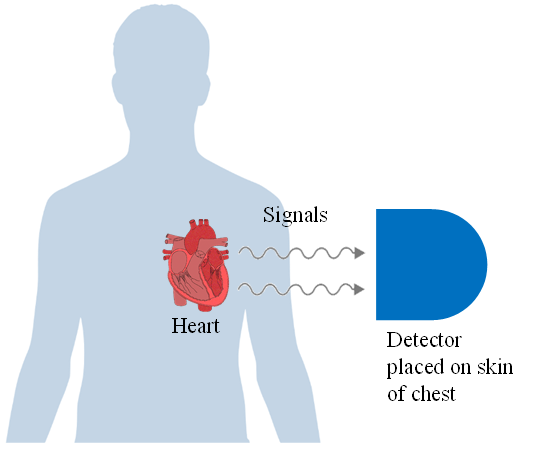}}
\caption{Basic diagram representing a person with a partial occlusion in their heart (CAD) and acoustic signals emanating from this to the detector surface (Clipart Library 2019).}
\label{diagexperi}
\end{figure}

Coronary Artery Disease (CAD) results from plaque deposit in a coronary artery. This reduces the amount of oxygenated blood from reaching the heart, thus early diagnosis is imperative in treating the disease and preventing death. Hence, non-invasive, accurate detection methods are of great interest. One of these methods identifies acoustic signals caused by partial occlusions, as the blood  flow in the artery is disturbed and causes oscillatory stresses on the artery wall. The deformations caused by the stresses can be detected at the chest surface. Figure \ref{diagexperi} is a diagram showing this concept. 

By using data gathered from simulating these surface signals, machine learning (ML) can be utilised to predict the source of the occlusion. This can be used by health professionals for diagnosis and more effective treatment. 

Preliminary experiments were conducted and it was found that the method used was time consuming. Hence, the investigation was limited and this motivates the use of ML in this report. ML is a set of tools for solving problems by understanding complex datasets and can learn and improve automatically.

The type of problem this report focuses on is a regression problem. ML algorithms can be used to predict the value of a continuous attribute. In this case, we wish to predict the value of the coordinates of the source disturbance ($x_c, y_c, z_c$) using the sensor readings ($u_1, u_2, u_3$). Thus, several ML algorithms were explored: Linear Regression (LR), XGBoost, Decision Tree (DT), Neural Network (NN), k-Nearest Neighbours (kNN), Random Forest (RF) and an ensemble model. 

The research challenge that this project aims to solve is: given some generated surface sensor readings can a ML model predict the source of this signal, and therefore detect the position of the occlusion? The overall aim of this project is to investigate whether ML can be used to solve this problem, as there is currently no existing system that does this. Furthermore, this project aims to find the best performing ML model for this task through experimenting with different algorithms and tuning methods.

\section{Related Work}
%%PAPER 1
Several studies were reviewed and analysed in order to better understand current methods and limitations of CAD diagnosis using ML. The earliest study from Hesser et al. (2020) developed ML algorithms in order to predict the location of an active source using the response from piezoelectric sensors. The data used for this was gathered by dropping a steel ball on an aluminium plate at different positions, and measuring the resulting elastic waves through sensors. Furthermore, numerical data was also gathered by performing 500 simulations of this experiment. 

Both the numerical and experimental data was used to train an Artificial Neural Network (ANN) and a Support Vector Machine (SVM) in order to predict the impact location of the steel ball based on the wave response of each sensor. The results showed that ML algorithms can precisely identify the location of a source. One evaluation method they used was the distance, in the $x$ and $y$ direction, between the source and the prediction of the ANN and SVM trained with both numerical and experimental data. 

The limitations of this study are that only two types of ML algorithms were investigated and a small dataset was used to train these models. The difference between the paper from Hesser et al. (2020) and this study is that the experimental setup only generated 2D numerical data. Specifically, the study reviewed was only able to locate the $x$ and $y$ coordinates of source whereas, in this study the $z$ coordinate will also be located. Furthermore, the study reviewed used a steel ball and a thin aluminium plate. However, in this study a generated occlusion embedded in a rectangular hexahedral domain will be used. Overall, this paper shows that it is possible to use ML algorithms to locate the coordinates of a source using data from sensors.

%%PAPER 2
The second study from Nasarian et al. (2020) used a dataset containing workplace and environmental features to train a ML algorithm. The main focus of this paper is the proposal of a new feature selection algorithm called heterogeneous hybrid feature selection (2HFS). The dataset used was the Nasarian CAD dataset and some workplace features included are office location of patients, noise exposure and pollutants. They also tested their method on other CAD datasets such as the Z-Alizadeh Sani dataset. 

The ML algorithms which were investigated are a Decision Tree (DT), Gaussian Naïve Bayes, Random Forest (RF), XGBoost, k-Nearest Neighbours (kNN) and Bernoulli Naïve Bayes. The results showed that that these workplace and environmental features have an impact on the detection of CAD. Specifically, the best ML algorithm in this study, XGBoost, achieved an accuracy of 81.23\%. 

The limitations of this study is that no neural networks (NN) were investigated and all the subjects in the Nasarian dataset were male, thus the results are not fully representative of the population. The difference between the paper from Nasarian et al. (2020) and this study is the type of data used. The study reviewed used categorical data whereas, in this paper numerical data in terms of signals will be used to train ML algorithms. Furthermore, the models in the reviewed study predicted whether a patient was likely to have CAD whereas, the model in this paper will predict the source of the occlusion which causes CAD. Overall, this study showed that feature selection during model training is important and has an effect on accuracy.

%%PAPER 3
The third study from Akella et al. (2021) investigated different ML algorithms in order to predict the likelihood of a person having CAD. Unlike previous studies beforehand, this study made their code used to build these ML models available publicly, so it can be used to assist healthcare workers in diagnosing CAD. The dataset utilised in this study is `the Cleveland dataset’ which consists of 14 variables obtained from 303 patients. Examples of variables in the dataset include age, blood cholesterol and chest pain. 

The ML algorithms they investigated were LR, regression tree, RF, SVM, kNN and NN. The results shows that NNs achieved the highest accuracy, 93\%, in predicting the presence of CAD. The limitation of this study is that the dataset used is quite small and quite old, it is from 1988, therefore the data may not be representative of the population today. Furthermore, some of the ML algorithms utilised are very similar to each other, e.g. DT and RF, therefore it would have been interesting to explore more varied algorithms and compare their results. Similarly to the previous paper, the difference between the paper from Akella et al. (2021) and this study is the type of data used and what the ML model predicts.

%%PAPER 4
The fourth paper from Doppala et al. (2022) aimed to improve the accuracy of ML methods for cardiovascular disease diagnosis by utilising an ensemble model. The data was obtained from the Mendeley Data Centre, IEEE DataPort and Cleveland dataset. This paper is different from previous studies as it combines four algorithms in order to create an ensemble model which achieves a higher accuracy than previous ML models. The ensemble model was a combination of naïve Bayes, RF, SVM and XGBoost. Each classifier made a prediction then, a ‘Voting Classifier’ selected the most popular outcome. 

The results showed that the proposed ensemble model achieves a greater accuracy than the benchmark classifiers (DT, RF, Naïve Bayes, Logistic Regression, SVM, Gradient Boosting, XGBoost). The limitations of this study are that NNs were not investigated. The difference between the paper from Doppala et al. (2022) and this study is that the dataset of the reviewed study consisted of categorical data such as age, gender, resting ECG etc. whereas, in this paper numerical data in terms of signals will be used to train a ML model. Overall, this paper showed how a voting classifier can be used in order to combine several different models together, which improves the accuracy of CAD diagnosis compared to benchmark classifiers.

\section{Methodology}
% Requirements capture / analysis – what your system should do
A computational model that simulates the sensor readings as a result of a source disturbance in the coronary artery will be utilised. The source disturbance is centred at a point $\boldsymbol{x}_c = (x_c,y_c,z_c)$ inside a rectangular hexahedral domain. It is of the form: 

\begin{equation} \boldsymbol{F}(\boldsymbol{x}) :=
A(1,1,1)^T \exp\left(
-\frac{\Vert\boldsymbol{x}-\boldsymbol{x}_c\Vert_{\mathbb{E}}^2}{\epsilon}\right)\end{equation}\label{force} where $A$ is a constant, and $\epsilon$ is small and positive to control the `local-ness' (Shaw and Greenwald 2022). 

\begin{figure}[!ht]
\centerline{\includegraphics[width=0.4\textwidth]{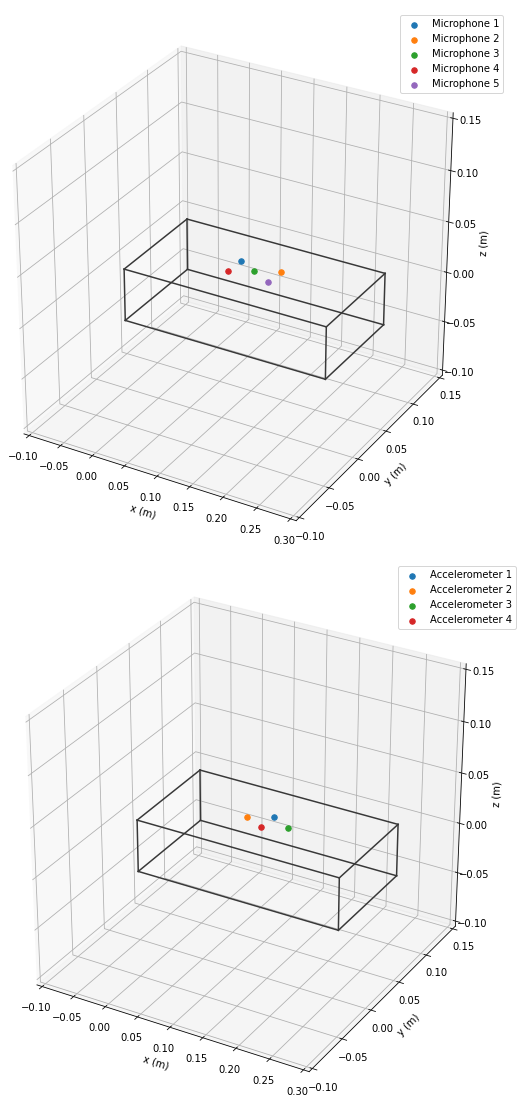}}
\caption{Basic diagrams of the experimental set up. The markers in the top diagram represent the locations of the microphones. The markers in the bottom diagram represent the locations of the accelerometers.}
\label{experiment}
\end{figure}

The body has a dimensions of $[0.0, 0.3]\times[-0.05, 0.05]\times[0.0, 0.05]$, in metres, along the $x$, $y$ and $z$ axes. The sensor readings measure the static displacement field as a response to this disturbance. The displacements $\boldsymbol{u} = (u_1, u_2, u_3)^T$ and other quantities, such as derivatives, are measured at sites on the surface using either five microphones or four accelerometers. The coordinates of the five microphones are: $(x,y)$ coordinates: $(0.12, 0.01)$, $(0.18, 0.01)$, $(0.15, 0.00)$, $(0.12,-0.01)$, $(0.18,-0.01)$. The coordinates of the four accelerometers are: $(x,y)$ coordinates: $(0.15, 0.01)$, $(0.12, 0.00)$, $(0.18, 0.00)$ and $(0.15,-0.01)$. These sensors are located on the top face at $z=0.05$. A schematic diagram of the experimental setup can be seen in Figure \ref{experiment}. In the simulation there is no intrinsic difference between the simulated microphone and accelerometer data, other than the site at which it is collected.

These sensor readings will then be used to train various ML algorithms to predict the source of the disturbance. The system should simulate data which is similar to real-life results, as the purpose of this research is to later use the ML model on patient data readings. Another requirement of the system is to generate many readings. This is because, ML algorithms require a large amount of data for training in order to generate good results when testing. Furthermore, the ML algorithms should perform just as well during deployment on unseen data as during training. 

% Design – how you went about your work
The data obtained contains $u_1$, $u_2$ and $u_3$ which represent the displacements in the $x$, $y$ and $z$ directions. Furthermore for each of these displacements, $\boldsymbol{u}_x$, $\boldsymbol{u}_y$ and $\boldsymbol{u}_z$ represent the derivative with respect to the $x$, $y$ and $z$ direction. These measurements are obtained for each of the 5000 randomly generated points $\boldsymbol{x}_c$. Also, these sets are replicated for two different narrowness of the source $\epsilon$ = 0.01 and $\epsilon$ = 0.001, for two different refinements of the finite element mesh 10 $\times$ 5 $\times$ 4 mesh and 20 $\times$ 10 $\times$ 8 mesh, and lastly for two different approximating polynomial degrees $r$ = 1 and $r$ = 2. Hence there are eight datasets with 5000 data points in each (Shaw and Greenwald 2022). In addition, for each run a randomly chosen source location, $\boldsymbol{x}_c = (x_c,y_c,z_c)$, was also obtained. 

Some pre-processing steps were taken before applying the ML algorithms. Firstly, for this investigation the three datasets $u_1$, $u_2$ and $u_3$ were combined with their derivatives into one singular dataset. This data was then split into a training and validation dataset, where 70\% of the data was used for training and 30\% was used for validation in order to tune the models. However, when implementing the NN the validation dataset is split again to create a testing dataset. So, the split for the NN is 70\% for training, 20\% for validation and 10\% for testing. This is because, NNs use the validation dataset while training therefore, an unseen dataset is needed when making predictions to avoid overfitting. After this, the singular dataset containing the $u_1$, $u_2$ and $u_3$ displacements and derivatives were normalised using the following equation: 

\begin{equation} z = \frac{x-\mu}{\sigma} \end{equation}\label{normal}
where $\mu$ is the mean and $\sigma$ is the standard deviation. This is done to ensure that inputs will be treated equally and have a similar range of amplitudes. 

The algorithms which will be investigated are: LR, DT, RF, XGBoost, kNN, NN and an ensemble model. Linear regression (LR) is one of the most simple ML algorithms. This is because it assumes that the underlying relationship between the independent and dependent variables is linear. Specifically for this study, multiple linear regression will be used as there are three predictors (the $x$, $y$ and $z$ coordinates of the source disturbance). The `LinearRegression' module from Scikit-Learn will be utilised (Pedregosa et al. 2011) and there are no parameters to be tuned. Overall, this algorithm will be implemented as a baseline in order to compare to more complex models. 

There are two key attributes of a Decision Trees (DT) which are nodes and branches. A node represents a subdivision of the DT and there are three types: the root node, internal node and leaf nodes. The root node is at depth 0 and represents a choice the algorithm must make which results in two mutually exclusive subsets of the DT (Song et al. 2015). An internal node represents splits in the tree between different choices. In a DT for regression these choices are based on values. Lastly, the leaf nodes represent the final result of the DT. Branches represent the pathway between different types of nodes based on the outcomes of the choices made by the nodes.

The `DecisionTreeRegressor' module from Scikit-Learn will be utilised (Pedregosa et al. 2011). This algorithm is trained using the structure of a tree, as described before, to predict a continuous output of the coordinates of the source disturbance. The feature which will be tuned in this model is `max\_depth' which is the number of layers in the DT. 

The Random Forest (RF) model is an ensemble of decision trees. The combination of these DTs for regression results in a very powerful model. This algorithm randomises the training samples in each instance of the decision tree, in order to reduce the risk of overfitting. The output of this model is the average prediction of each individual tree. The `RandomForestRegressor' module from Scikit-Learn will be utilised (Pedregosa et al. 2011). The features which will be tuned in this model are `n\_estimators' which is the maximum number of estimators (trees) used to train and `max\_depth'. 

XGBoost stands for Extreme Gradient Boosting. The reason for using a ``Boosted" decision tree for regression is because, a DT is known as a weak learner which is a classifier that performs only slightly better than random guessing, and they can be easily over-trained (Géron 2017). However, `boosting' makes the DT become a strong learner. A stronger learner is a classifier which achieves high accuracy (Géron 2017). Boosting is an example of a re-weighting technique where mis-classified events are re-weighted from one iteration of training to the next. Therefore, subsequent training iterations focus on correcting mis-classified events. The `XGBRegressor' class will be utilised (Chen et al. 2016). The features which will be tuned in this model are `n\_estimators', `max\_depth' and `learning\_rate' which is the step size for the gradient boosting algorithm. 

The k-Nearest Neighbours (kNN) model is a non-parametric approach that does not have any explicit decision boundaries, rather a sample is assigned the label of the most similar training sample. It does this by calculating the Euclidean distance to all the training samples, then selects the `k' closest samples (neighbours) and assigns them a label. The `KNeighborsRegressor' module from Scikit-Learn will be utilised (Pedregosa et al. 2011). The features which will be tuned in this model are `n\_neighbors', which represents the number of neighbours, `k'. Also, the parameter `weights', which affects the influence samples have in each neighbourhood.  

The neural network (NN) in this report consists of single layers connected sequentially. The data passes through the input layer, hidden layers and lastly an output layer. Each perceptron within a layer is defined by weights, activation functions and biases. NNs also have a loss function which measures the performance of the NN by comparing the desired output to the actual output, and calculates the output error (Géron 2017). For this investigation, the `Mean Squared Error` loss function will be used. TensorFlow specifically with Keras will also be used to create a NN in Python.

An ensemble model is a combination of different ML algorithms with a `Voting Regressor' and is the approach the study from Doppala et al. (2022) also investigated. They found that that accuracy of their model improved using this method. A combination of all models was tested however, the ensemble model performed better when using just RF and kNN to train. This is because, ensemble learning works best when training with a diverse range of models, as this varies the types of errors and thus improves the ensemble's accuracy (Géron 2017). The purpose of the `Voting Regressor' is to average the individual predictions to output a final prediction. The `VotingRegressor` module from Scikit-Learn will be utilised (Pedregosa et al. 2011). The parameters of the RF and kNN will be tuned separately, and the best values for each parameter will be used in the ensemble model.

% Implementation – practical techniques, problems, solutions
% Testing and/or evaluation – how well your solution worked

To tune these models a module from Scikit-Learn will be utilised called `GridSearchCV'. This technique searches for the optimal value to use for each specified parameter. Therefore, instead of manually tuning each model, `GridSearchCV' does this automatically. The input for this function are the parameters that require tuning for each model. This is typically `n\_estimators', `max\_depth' and `learning\_rate'. The algorithm then computes all the possible combinations of these values, and comes to a conclusion on which is the best combination using cross-validation (Géron 2017). 

In order to implement all of these different ML algorithms, Python will be used with Jupyter Notebooks. The evaluation techniques used will be the Mean Squared Error (MSE):

\begin{equation} E_{MSE} = \frac{1}{N} \sum^{N}_{i=1} (y_i - \hat{y}_i)^2 \end{equation} \label{mse} where, $y_i$ is the continuous true label of the sample $i$, $\hat{y}_i$ is the predicted label for the sample and $N$ is the number of samples in the dataset. Furthermore, the average distance between the prediction on the validation dataset and the ground truth of the source disturbance will also be calculated using the Euclidean distance: 

\begin{equation} d(y_i,\hat{y}_i) = \sqrt{\sum^{N}_{i=1} (y_i - \hat{y}_i)^2}\end{equation}\label{dist} where, $y_i$ and $\hat{y}_i$ represent the ground truth and prediction respectively and $N$ represents the number of samples. This is a similar evaluation metric as used in Hesser et al. (2020). Lastly, 3D plots were created with the same geometry as the hexahedral domain to visualise the difference between the ground truth and prediction of the source disturbance. 

These evaluation measures will also be used to assess whether the model is being overfitted or underfitted. The model will display overfitting if it has a very low training MSE but a high validation MSE. This occurs when the model is memorising irrelevant details of the training samples, and thus does not perform well during validation. A model will display underfitting if the training and validation MSEs are high. In this case, the model is unable to identify the underlying pattern within the data.

\section{Results}
%Plots of algorithms, explain the results of each algorithm, MSE, 
Several tests were conducted in order to investigate which ML technique achieved the best results. This included using different subsets of the data depending on the mesh resolution, $\epsilon$ value, polynomial degree and whether it contained microphone data, accelerometer data or both. Firstly, only one single file corresponding to $\epsilon=0.01$, polynomial degree $r=1$ and 10 $\times$ 5 $\times$ 4 mesh and only microphone data was used to train each model. 

Table \ref{Table1} shows the validation MSEs after training and tuning the models as well as the respective MSEs of the $x$, $y$ and $z$ coordinates. The validation MSE was calculated using equation \ref{mse}, where $y_i$ is from the validation dataset and $\hat{y}_i$ is the models' prediction on the validation dataset. The MSEs of the $x$, $y$ and $z$ coordinates was calculated using a similar methodology however, the prediction for each individual coordinate was used so three different MSEs for each coordinate was computed.

\begin{table}[!ht]
\begin{tabular}{lclll}
\hline
\multicolumn{1}{c}{\multirow{2}{*}{\textbf{Model}}} & \multirow{2}{*}{\textbf{\begin{tabular}[c]{@{}c@{}}Validation MSE\\  ($\times10^{-5}$)\end{tabular}}} & \multicolumn{3}{c}{\multirow{2}{*}{\textbf{\begin{tabular}[c]{@{}c@{}}MSE \\ ($\times10^{-5}$)\end{tabular}}}} \\
\multicolumn{1}{c}{} &  & \multicolumn{3}{c}{} \\ \hline
\multirow{2}{*}{} & \multicolumn{1}{l}{\multirow{2}{*}{}} & \multicolumn{1}{c}{\multirow{2}{*}{x}} & \multicolumn{1}{c}{\multirow{2}{*}{y}} & \multicolumn{1}{c}{\multirow{2}{*}{z}} \\
 & \multicolumn{1}{l}{} & \multicolumn{1}{c}{} & \multicolumn{1}{c}{} & \multicolumn{1}{c}{} \\
\multirow{2}{*}{Linear Regression} & \multirow{2}{*}{27.1} & \multirow{2}{*}{66.1} & \multirow{2}{*}{9.66} & \multirow{2}{*}{5.53} \\
 &  &  &  &  \\
\multirow{2}{*}{XGBoost} & \multirow{2}{*}{6.77} & \multirow{2}{*}{8.02} & \multirow{2}{*}{3.73} & \multirow{2}{*}{8.56} \\
 &  &  &  &  \\
\multirow{2}{*}{Decision Tree} & \multirow{2}{*}{5.86} & \multirow{2}{*}{8.10} & \multirow{2}{*}{2.49} & \multirow{2}{*}{6.99} \\
 &  &  &  &  \\
\multirow{2}{*}{Neural Network} & \multirow{2}{*}{3.14} & \multirow{2}{*}{1450} & \multirow{2}{*}{172} & \multirow{2}{*}{46.6} \\
 &  &  &  &  \\
\multirow{2}{*}{kNN} & \multirow{2}{*}{1.62} & \multirow{2}{*}{1.01} & \multirow{2}{*}{0.57} & \multirow{2}{*}{3.27} \\
 &  &  &  &  \\
\multirow{2}{*}{Random Forest} & \multirow{2}{*}{2.53} & \multirow{2}{*}{3.72} & \multirow{2}{*}{0.72} & \multirow{2}{*}{3.15} \\
 &  &  &  &  \\
\multirow{2}{*}{Ensemble} & \multirow{2}{*}{1.69} & \multirow{2}{*}{1.85} & \multirow{2}{*}{0.37} & \multirow{2}{*}{2.84} \\
 &  &  &  &  \\ \hline
\end{tabular}
\centering
\caption{The validation MSEs and the MSEs of the $x$, $y$ and $z$ coordinates for the different models using only one single file corresponding to $\epsilon=0.01$, polynomial degree $r=1$ and 10 $\times$ 5 $\times$ 4 mesh and only microphone data.}
\label{Table1}
\end{table}

\begin{figure}[!ht]
\centering
\centerline{\includegraphics[width=0.45\textwidth]{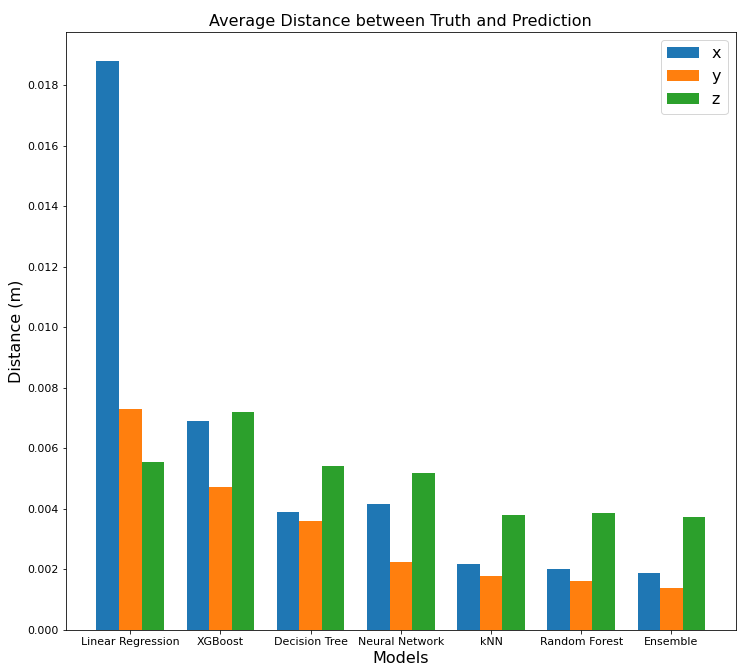}}
\caption{The average distance between the prediction on the validation dataset and the ground truth for all models and coordinates. Only one single file corresponding to $\epsilon=0.01$, polynomial degree $r=1$ and 10 $\times$ 5 $\times$ 4 mesh and only microphone data was used.}
\label{14avgdis}
\end{figure}

These results show that the three best performing models are kNN, RF and an ensemble model, as both their validation MSE and the MSEs of each coordinate are lower than the other models. To further investigate this, the average distance between the prediction and the ground truth was plotted for all coordinates and models. This can be seen in Figure \ref{14avgdis}.

This figure shows that the ensemble model was the most accurate in predicting the coordinates of the source disturbance. In addition, the results show that the $y$-coordinate of this disturbance was the best predicted coordinate for all models except LR. Whereas, the $z$-coordinate was the most difficult coordinate for the kNN, RF and ensemble model to predict.

\begin{figure}[!ht]
\centering
\centerline{\includegraphics[width=0.4\textwidth]{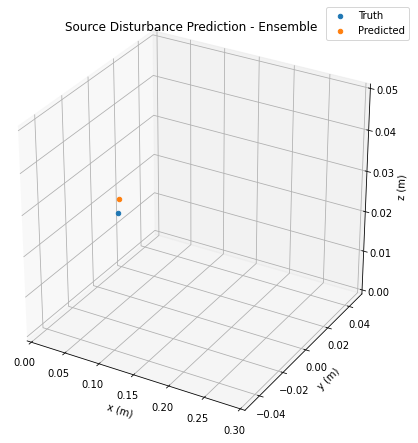}}
\caption{A 3D plot comparing the ground truth to the prediction for the ensemble model for a single example. Only one file corresponding to $\epsilon=0.01$, polynomial degree $r=1$ and 10 $\times$ 5 $\times$ 4 mesh and only microphone data was used. Figure is not to scale.}
\label{143dplot}
\end{figure}

A 3D plot showing the truth in comparison to the predicted values for the ensemble model can be seen in Figure \ref{143dplot}. This plot shows that the prediction made by the ensemble model is near to the ground truth, with only a small deviation. Specifically, the coordinate with the largest deviation from the truth and prediction is the $z$-coordinate, which is averaged to be 3.7mm. In comparison, the $y$-coordinate has the smallest average difference between the truth and prediction of 1.4mm.   

\begin{figure}[!ht]
\centering
\centerline{\includegraphics[width=0.5\textwidth]{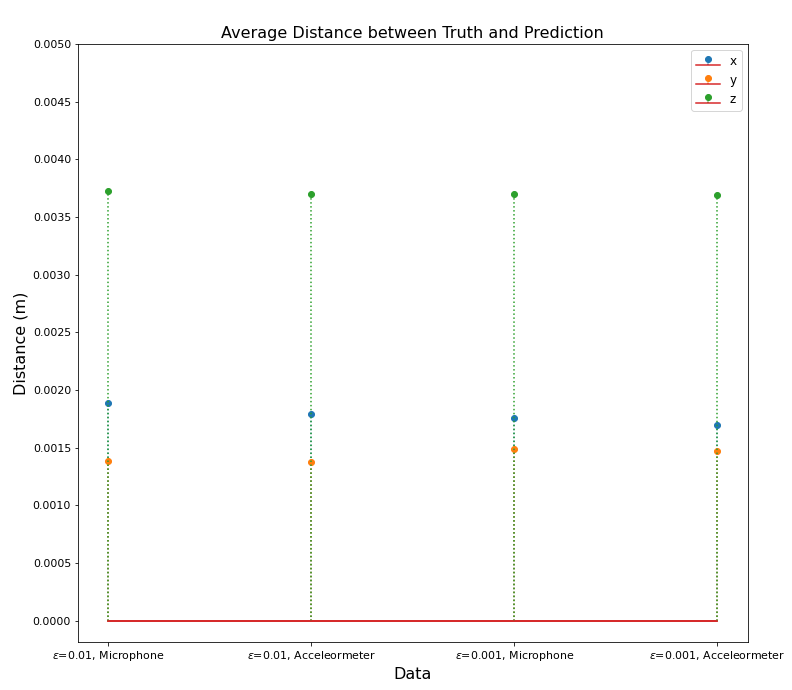}}
\caption{The average distance between the prediction using the ensemble model and the ground truth, for each coordinate. Using only one file corresponding to polynomial degree $r=1$ and 10 $\times$ 5 $\times$ 4 mesh for $\epsilon=0.01$ and $\epsilon=0.001$, with microphone and accelerometer data.}
\label{1415micacc}
\end{figure}

Similarly to before, one single file corresponding to $\epsilon=0.01$ polynomial degree $r=1$ and 10 $\times$ 5 $\times$ 4 mesh was used however, only accelerometer data was used to train the models. There was no major difference in the results when comparing microphone and accelerometer data. A comparison between the microphone and accelerometer data for this configuration, as well as as for $\epsilon=0.001$ polynomial degree $r=1$ and 10 $\times$ 5 $\times$ 4 mesh can be seen in Figure \ref{1415micacc}.

It shows that the $x$-coordinate prediction was slightly better, and the $y$-coordinate prediction was slightly worse for $\epsilon = 0.001$. Furthermore, it shows that for both these configurations, the $y$-coordinate error was the smallest and the $z$-coordinate error was the largest. This shows that the narrowness of the source ($\epsilon$) is almost irrelevant in predicting the location of the source disturbance.

Secondly, two models were created using all data either corresponding to $\epsilon=0.01$ or $\epsilon=0.001$. This contained data using polynomial degree $r=1$ and $r=2$, as well as 10 $\times$ 5 $\times$ 4 mesh and 20 $\times$ 10 $\times$ 8 mesh. Furthermore, two subsets using this data were carried out, first using only microphone data and then using only accelerometer data. 

The parameters used for the RF model were, `n estimators’: 800 and  `max depth’: 25. The parameters used for the kNN model were, `n estimators’: 4 and `weights': `distance'. Table \ref{Table3} shows the MSEs of the ensemble model trained using data corresponding to $\epsilon=0.01$ and $\epsilon=0.001$ with microphone or accelerometer data. Comparing these results to the first test, where only one file corresponding to  $\epsilon=0.01$, polynomial degree $r=1$ and 10 $\times$ 5 $\times$ 4 mesh was used, the $z$-coordinate error has increased.  However, the $x$ and $y$ coordinate errors of the ensemble model decreased when using these models.

\begin{table}[!ht]
\begin{tabular}{clclll}
\hline
\multicolumn{1}{l}{\multirow{2}{*}{\textbf{$\epsilon$ Value}}} & \multicolumn{1}{c}{\multirow{2}{*}{\textbf{Type of Data}}} & \multirow{2}{*}{\textbf{\begin{tabular}[c]{@{}c@{}}Validation MSE\\  ($\times10^{-5}$)\end{tabular}}} & \multicolumn{3}{c}{\multirow{2}{*}{\textbf{\begin{tabular}[c]{@{}c@{}}MSE \\ ($\times10^{-5}$)\end{tabular}}}} \\
\multicolumn{1}{l}{} & \multicolumn{1}{c}{} &  & \multicolumn{3}{c}{} \\ \hline
\multicolumn{1}{l}{\multirow{2}{*}{}} & \multirow{2}{*}{} & \multicolumn{1}{l}{\multirow{2}{*}{}} & \multicolumn{1}{c}{\multirow{2}{*}{x}} & \multicolumn{1}{c}{\multirow{2}{*}{y}} & \multicolumn{1}{c}{\multirow{2}{*}{z}} \\
\multicolumn{1}{l}{} &  & \multicolumn{1}{l}{} & \multicolumn{1}{c}{} & \multicolumn{1}{c}{} & \multicolumn{1}{c}{} \\
\multirow{4}{*}{0.01} & \multirow{2}{*}{Microphone} & \multirow{2}{*}{1.44} & \multirow{2}{*}{0.49} & \multirow{2}{*}{0.29} & \multirow{2}{*}{3.55} \\
 &  &  &  &  &  \\
 & \multirow{2}{*}{Accelerometer} & \multirow{2}{*}{1.33} & \multirow{2}{*}{0.45} & \multirow{2}{*}{0.28} & \multirow{2}{*}{3.26} \\
 &  &  &  &  &  \\
\multirow{4}{*}{0.001} & \multirow{2}{*}{Microphone} & \multirow{2}{*}{1.53} & \multirow{2}{*}{0.59} & \multirow{2}{*}{0.29} & \multirow{2}{*}{3.71} \\
 &  &  &  &  &  \\
 & \multirow{2}{*}{Accelerometer} & \multirow{2}{*}{1.36} & \multirow{2}{*}{0.43} & \multirow{2}{*}{0.26} & \multirow{2}{*}{3.40} \\
 &  &  &  &  &  \\ \hline
\end{tabular}
\centering
\caption{The validation MSEs and the MSEs of the x, y and z coordinates for the ensemble model using all data corresponding to $\epsilon=0.01$ or $\epsilon=0.001$ with microphone or accelerometer data.}
\label{Table3}
\end{table}

\begin{figure}[!ht]
\centering
\centerline{\includegraphics[width=0.5\textwidth]{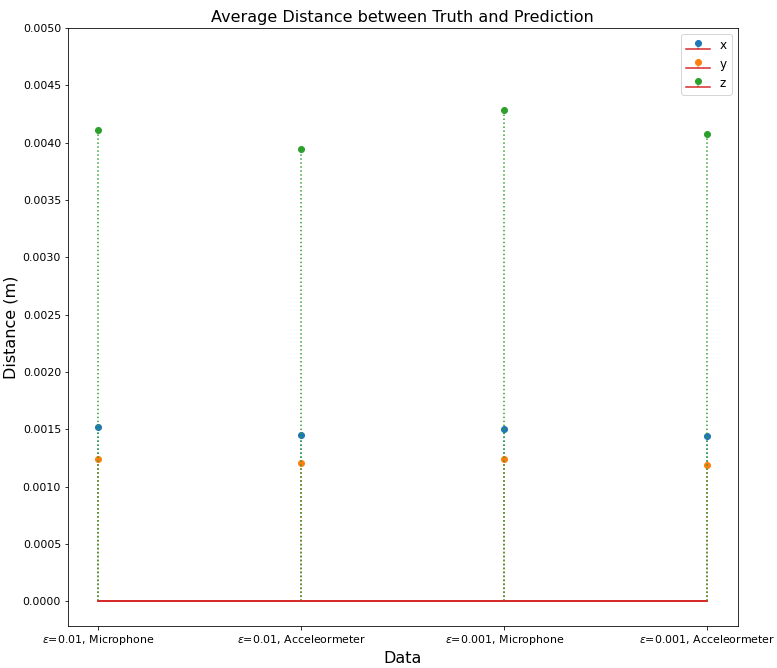}}
\caption{The average distance between the prediction using the ensemble model and the ground truth, for each coordinate. Using all data corresponding to either $\epsilon=0.01$ or $\epsilon=0.001$, with either microphone or accelerometer data.}
\label{1617micacc}
\end{figure}

To investigate this further, Figure \ref{1617micacc} shows the average distance between the ground truth and prediction for the two models using all data corresponding to $\epsilon=0.01$ or $\epsilon=0.001$, with either microphone or accelerometer data. These results show that regardless of the type of data, the ensemble model's prediction is very similar. This was also seen in Figure \ref{1415micacc}. This is to be expected as the simulated microphone and accelerometer data are fundamentally the same.

Further comparing this to the previous test using only one file, the overall average distance between the truth and predicted $x$-coordinate decreased, $y$-coordinate decreased and $z$-coordinate increased. This shows that, alongside the MSEs, the $x$ and $y$ coordinate prediction of the ensemble model improved when using all data corresponding to either $\epsilon=0.01$ or $\epsilon=0.001$. This is because, the more data the ML algorithms have, the more samples can be used for training. Therefore, the models can make more accurate predictions. 

% This can also be visualised with a 3D plot. Figure \ref{163dplot} shows these results using the microphone data. Comparing this to Figure \ref{143dplot}, it shows that using this configuration achieved more accurate results as there is only a small deviation between the ground truth and prediction. 

% \begin{figure}[!ht]
% \centering
% \centerline{\includegraphics[width=0.4\textwidth]{Dissertation_Paper_Template - LATEX/Figures/6.5_ensemble_results_test_2.png}}
% \caption{A 3D plot comparing the ground truth to the prediction on the validation dataset for the ensemble model for a single example. Using all data corresponding to $\epsilon=0.01$ with microphone data.}
% \label{163dplot}
% \end{figure}

Thirdly, two models were created using all data either corresponding to $\epsilon=0.01$ or $\epsilon=0.001$. This contained data using polynomial degree $r=1$ and $r=2$, as well as 10 $\times$ 5 $\times$ 4 mesh and 20 $\times$ 10 $\times$ 8 mesh. However, unlike the previous test, both microphone and accelerometer data were combined. This resulted in all models, including the ensemble model, having similar results. They were all unable to accurately predict the source disturbance location. Therefore, the ML algorithms underfitted as they were unable to learn the underlying pattern of the data.

Lastly, two universal models were created using all data corresponding to both $\epsilon=0.01$ and $\epsilon=0.001$. This contained data using polynomial degree $r=1$ and $r=2$, as well as 10 $\times$ 5 $\times$ 4 mesh and 20 $\times$ 10 $\times$ 8 mesh. Furthermore, either microphone or accelerometer data was used. These two types of data were investigated separately due to the results from the previous test.

The parameters used for the RF model were, `n estimators’: 800 and  `max depth’: 25. The parameters used for the kNN model were, `n estimators’: 3 and `weights': `distance'. Table \ref{Table4} shows that MSEs of the ensemble model using this configuration. Comparing these results to the second test, where the two $\epsilon$ values were separated, the MSEs for all the coordinates have decreased. 

\begin{table}[!h]
\begin{tabular}{lclll}
\hline
\multicolumn{1}{c}{\multirow{2}{*}{\textbf{Type of Data}}} & \multirow{2}{*}{\textbf{\begin{tabular}[c]{@{}c@{}}Validation MSE\\  ($\times10^{-5}$)\end{tabular}}} & \multicolumn{3}{c}{\multirow{2}{*}{\textbf{\begin{tabular}[c]{@{}c@{}}MSE \\ ($\times10^{-5}$)\end{tabular}}}} \\
\multicolumn{1}{c}{} &  & \multicolumn{3}{c}{} \\ \hline
\multirow{2}{*}{} & \multicolumn{1}{l}{\multirow{2}{*}{}} & \multicolumn{1}{c}{\multirow{2}{*}{x}} & \multicolumn{1}{c}{\multirow{2}{*}{y}} & \multicolumn{1}{c}{\multirow{2}{*}{z}} \\
 & \multicolumn{1}{l}{} & \multicolumn{1}{c}{} & \multicolumn{1}{c}{} & \multicolumn{1}{c}{} \\
\multirow{2}{*}{Microphone} & \multirow{2}{*}{1.00} & \multirow{2}{*}{0.31} & \multirow{2}{*}{0.18} & \multirow{2}{*}{2.53} \\
 &  &  &  &  \\
\multirow{2}{*}{Accelerometer} & \multirow{2}{*}{0.92} & \multirow{2}{*}{0.27} & \multirow{2}{*}{0.17} & \multirow{2}{*}{2.31} \\
 &  &  &  &  \\ \hline
\end{tabular}
\centering
\caption{The validation MSEs and the MSEs of the x, y and z coordinates for the ensemble model using all data combining $\epsilon=0.01$ and $\epsilon=0.001$, with microphone or accelerometer data.}
\label{Table4}
\end{table}

\begin{figure}[!ht]
\centering
\centerline{\includegraphics[width=0.5\textwidth]{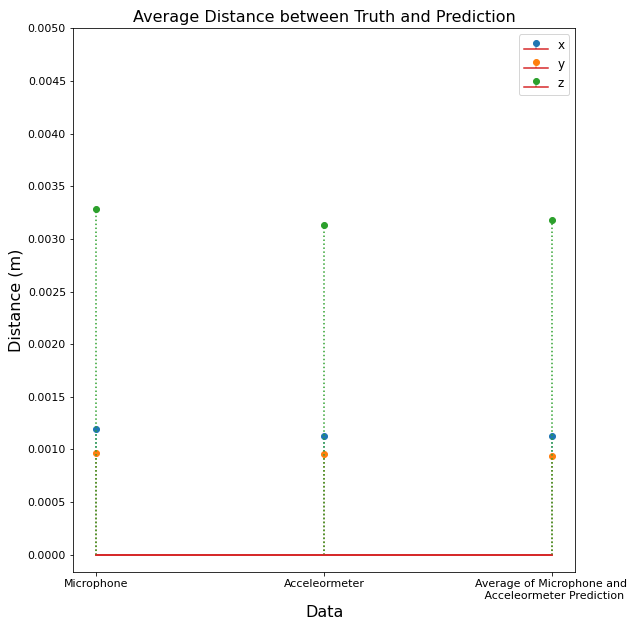}}
\caption{The average distance between the prediction using the ensemble model and the ground truth, for each coordinate. Using data corresponding to $\epsilon=0.01$ and $\epsilon=0.001$ with microphone or accelerometer data and an average of the microphone and accelerometer predictions.}
\label{2223micacc}
\end{figure}

In addition, Figure \ref{2223micacc} shows the average distance between the ground truth and prediction for the universal models for microphone or accelerometer data. Comparing these results to the previous tests, the overall average distance between the ground truth and predicted coordinates has decreased, particularly for the $x$ and $y$ coordinates. This shows that, alongside the MSEs, the source disturbance prediction of the ensemble model improved when using all data corresponding to a combination of $\epsilon=0.01$ and $\epsilon=0.001$. This further consolidates that the value of $\epsilon$ does not impede the ML models' prediction.

\begin{figure}[!ht]
\centering
\centerline{\includegraphics[width=0.4\textwidth]{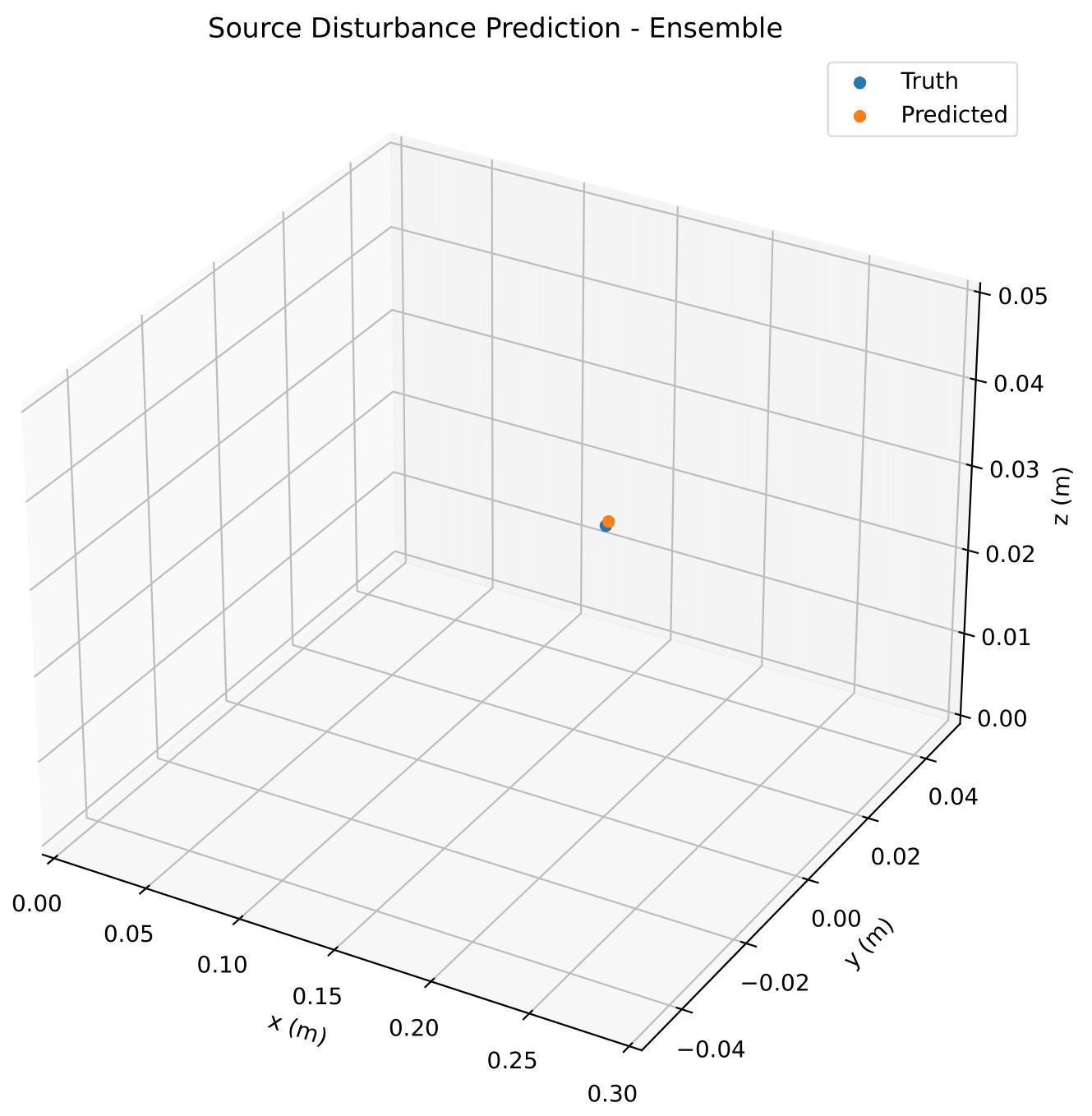}}
\caption{A 3D plot comparing the ground truth to the prediction on the validation dataset for the ensemble model for a single example. Using a combination of data corresponding to $\epsilon=0.01$ and $\epsilon=0.001$ with the average of the microphone and accelerometer prediction. Figure is not to scale.}
\label{24dplot}
\end{figure}

An extension to this test was done by combining the results from each universal model, either using microphone or accelerometer data, together. This was done by averaging the two predictions to create a final prediction. Then, the average distance between the ground truth and this final prediction was calculated and the results can be seen in Figure \ref{2223micacc}. It shows that the $x$ and $y$ coordinate prediction improved however, the $z$-coordinate prediction did not improve compared to when using just the accelerometer data. This can also be visualised with a 3D plot. Figure \ref{24dplot} shows the ensemble model's results using the average of the microphone and accelerometer prediction. Comparing this to Figure \ref{143dplot}, it shows an even smaller deviation from the ground truth and prediction. 

The $z$-coordinate prediction may have not improved as the sensors are less sensitive to signals along the $z$ axis. This may be due to the locations of the microphones and accelerometers. This can be seen in Figure \ref{experiment} of the experimental setup. It shows that the sensors are more spread out along the $x$ and $y$ axis compared to the $z$ axis.

\begin{figure}[!ht]
\centering
\centerline{\includegraphics[width=0.5\textwidth]{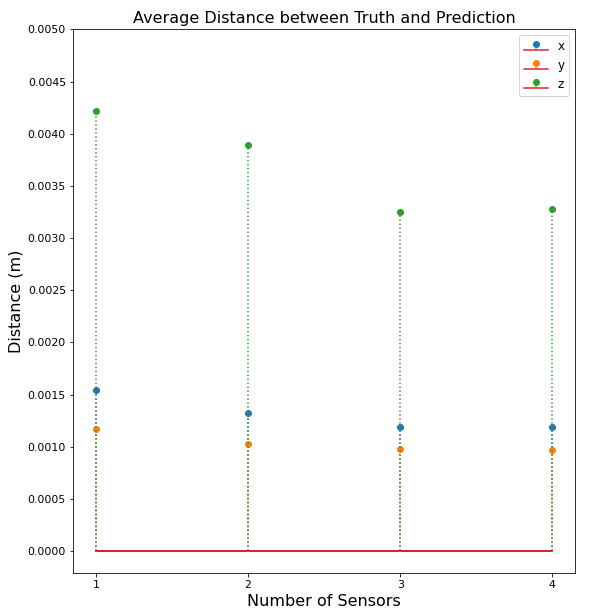}}
\caption{The average distance between the prediction using the ensemble model and the ground truth, for each coordinate. Using data corresponding to $\epsilon=0.01$ and $\epsilon=0.001$ with microphone data and a different number of sensors.}
\label{micsensors}
\end{figure}

A further test was conducted in order to investigate the effect of the number of sensors used. This was done by using the previous universal model with microphone data. However, the datasets utilised to train and validate the model correspond to either using one, two or three sensors. The locations of these sensors can be seen in Figure \ref{experiment} and they correspond to the results seen in Figure \ref{micsensors}. These results shows that the more sensors used in the experimental setup, the more accurate the models' prediction is. However, this is only up until a certain point as the results from using three and four sensors are very similar. Therefore, the positions of the sensors may also affect their sensitivity to signals and thus, the model's ability to predict the source position. 

\section{Discussion/Conclusion}
% This should be a critical analysis of your work and an honest appraisal of the achievements of your project.
In conclusion, simulated sensor readings were used to train ML models to predict the source location of this signal, and thus detect the position of the occlusion. Several regression ML algorithms were explored, as well as experimenting with different configurations of the data. In the end, it was found that ML models can predict, with high accuracy, the location of the source as a result of the sensor readings. 

Overall, there are several outcomes to this work. Firstly, the best performing ML algorithm was the ensemble model. This ensemble model was a combination of the kNN and Random Forest algorithms. 

Secondly, the best performing configuration was using all data corresponding to a combination of $\epsilon=0.01$ and $\epsilon=0.001$ with either microphone or accelerometer data. Furthermore, the average of these two predictions, to create a final estimate, resulted in the best performing model. The average MSE of the $x$, $y$ and $z$ coordinates were 2.9$\micro$m$^2$, 1.8$\micro$m$^2$ and 0.024mm$^2$ respectively as seen in Table \ref{Table4}. The average distance between the ground truth and prediction for the $x$, $y$ and $z$ coordinates were 1.1mm, 0.93mm, 3.2mm respectively as seen in Figure \ref{2223micacc}.

Thirdly, the results show that the narrowness of the source ($\epsilon$) does not greatly affect the models' prediction. This may be useful when building a non-invasive detection method for future real-life scenarios.

Lastly, the results showed that nearly all models predicted the $y$-coordinate of the source disturbance the most accurately. While the $z$-coordinate was the most inaccurate. A possible method to improve these results would be to spread the sensors along the $z$-axis more. However, this is limited to some extent as the sensors are confined to the skin surface, so the z-distance of the sensors cannot vary too much. Also, a further investigation is needed in order to see if the sites at which the microphones and accelerometers are located improves the prediction of the $z$-coordinate. This is critical in determining the optimal sensor positions in the real device.

\section{Future work}
Future work for this investigation would be to add noise to the simulated data. This is because, real-life readings are not perfect and often have background noise. Also, using time-dependent data would also be beneficial. These additions would simulate these situations more accurately and help improve the ML models' performance. 

However, testing the ensemble model developed on real-life data would provide the most meaningful results. This is because, the model's performance can be assessed and improved using the same type of data it will use during deployment.

All data used in this investigation was numerically generated. However, the results from this work are intended, in the future, to be used on real-life CAD patient data. Therefore, legal, social, ethical and sustainability issues such as privacy need to be considered when applying these ML models to personal data.

\vspace{12pt}

\end{document}